\documentclass[%
 reprint,
 amsmath,amssymb,
 aps,
 prl,
 longbibliography,
 lengthcheck,%
]{revtex4-1}

\usepackage{graphicx}
\usepackage{dcolumn}
\usepackage{bm}
\usepackage{hyperref}

\begin{document}

\preprint{APS/123-QED}

\title{Dipole-dipole interaction and polarization mode in BEC
}

\author{P. A. Andreev}%
\email{andreevpa@physics.msu.ru}
\author{L. S. Kuzmenkov}%
\email{lsk@phys.msu.ru} \affiliation{
 Faculty of Physics, Moscow State
University, Moscow, Russian Federation.}





\date{\today}

\begin{abstract}
We propose the construction of a set of quantum hydrodynamics
equations for the Bose-Einstein condensate (BEC) where atoms have
electric dipole moment (EDM). The contribution of the
dipole-dipole interactions (DDI) to the Euler equation is
estimated. Quantum equations for the evolution of medium
polarization are constructed for the first time. The mathematical
method we developed allows studying the effects of interactions on
the evolution of polarization. The developed method may be applied
to various physical systems in which dynamics is affected by DDI.
A problem of elementary excitations in BEC, either affected or not
affected by the uniform external electric field, is addressed
using our method. We show that the evolution of polarization in BEC leads
to the formation of a novel type of elementary excitations.
Also, we consider the process of wave generation in polarized BEC
by means of monoenergetic beam of neutral polarized particles. We
compute the possibilities of the generation of Bogoliubov's modes
and polarization modes by the dipole beam.
\end{abstract}

\maketitle


\section{\label{sec:level1}I. Introduction}
Since obtaining the Bose-Einstein condensate (BEC) in experiment with pairs of alkaline metal
atoms, theoretical and experimental investigation of linear waves
and nonlinear structures in it have been performed. The interest
to the spinor BEC ~\cite{Szankowski PRL 10,Cherng PRL 09,Lahaye
RPP 09}, the effect of magnetic moment on the BEC evolution
~\cite{Lahaye RPP 09,dynamic of magn moments,GP for magn,Lahaye
Nature} and the influence of electrical polarization of atoms on
the dynamic processes in BEC ~\cite{Lahaye RPP 09,Ticknor PRL
11,polarized BEC first step} is rising in recent years.

Many processes in quantum systems are determined by the dynamics
and dispersion of elementary excitations (EE) ~\cite{Griffin
book}. A law of dispersion of EE in degenerate dilute Bose gas was
obtained by Bogoliubov in 1947 ~\cite{Bogoliubov
N.1947,L.P.Pitaevskii RMP 99}. Later, many authors studied the
change of the Bogoliubov's mode which arise when interactions are
more carefully in ~\cite{Lin waves interact,Andreev PRA08}, a
geometry of the system are complex ~\cite{Lin waves geom,mass dep
of waves in BEC}. In article ~\cite{Ticknor PRL 11} authors
studied the influence of electric dipole moment (EDM) dynamics on
dispersion of EE in BEC. The contribution of polarization in
dispersion law of Bogoliubov's mode was obtained in ~\cite{Ticknor
PRL 11}. The polarization waves in low dimensional and multy-layer
systems of conductors and dielectrics are considered in the papers
~\cite{Andreev DSS 10,Qiuzi Li PRB 11}. Consequently, in BEC of
atoms with EDM we expect the existence of polarizations wave;
along with Bogoliubov's mode. It is particularly important to take
into account the dynamics of EDM in investigations of exciton BEC
in semiconductors. The method of QHD ~\cite{MaksimovTMP 1999} was
used in investigations of BEC ~\cite{Andreev PRA08} and
polarization waves in conductors and semiconductors ~\cite{Andreev
DSS 10}. Therefore, the method of QHD could be used to consider
the possibility of arising polarization waves in BEC and
calculating of the dispersion law of these waves.

For investigation of BEC, in system of particles with EDM or
magnetic moments, various theoretical methods are used. The model
equations, proposed for the description of spinor and polarized
BEC (such as the non-linear Schr\"{o}dinger equation (NLSE)
~\cite{Szankowski PRL 10,Lahaye RPP 09,GP for magn,Ticknor PRL
11}) are generalizations of the Gross-Pitaevkii equation
~\cite{L.P.Pitaevskii RMP 99}.

Other methods were applied for investigation of the polarized BEC
and other systems of particles, with EDM, along with NLSE. A
hydrodynamic formulation of the Hartree-Fock theory for particles
with significant EDM is considered in ~\cite{Lima PRA 10}. In
paper ~\cite{Lima PRA 10} the Euler-type equation was derived,
from the evolution of the density matrix. The EDM dynamics in
dimer Mott insulators causes the rise of the low-frequency mode
~\cite{Gomi PRB 10}. A two-body quantum problem for polarized
molecules is analyzed in ~\cite{Quemener PRA 11}. The existence of
states with spontaneous interlayer coherence has been predicted in
~\cite{Lutchyn PRA 10} in systems of polar molecules.
Calculations, in ~\cite{Lutchyn PRA 10}, were based upon the
secondary quantization approach, where the Hamiltonian accounts
for the molecular rotation and dipole-dipole interactions.
Superfluidity anisotropy of polarized fermion systems was shown in
~\cite{Liao PRA 10} and their thermodynamic and correlation
properties were investigated ~\cite{Baillie PRA 10}. The effect of
EDM in a system of cooled neutral atoms which are used for quantum
computing and quantum memory devices was analyzed in
~\cite{Gillen-Christandl PRA 11}.

The characteristic property of BEC in a system of excitons
inducing in semiconductors ~\cite{Deng RMP 10} it is a significant
value of EDM of excitons. This leads to strong interaction of
excitons with external electric field and emergence of the
collective DDI in exciton systems. QHD method may be applied to
such systems along with quantum kinetics based on the
nonequilibrium Green functions ~\cite{Haug book 08} or density
matrix equations ~\cite{Kuhn book 97}.

Notable success has been reached in the Bose condensation of dense
gases ~\cite{Vogl Nature 09,Sheik-Bahae NP 09}. Consequently, we
need the detailed account of short range interaction in
investigations of dynamics of EDM. In this work we account the
short range interaction (SRI) up to the third order on interaction
radius (TOIR) ~\cite{Andreev PRA08}. This approximation leads to
nonlocality of SRI ~\cite{Andreev PRA08,Rosanov and Braaten}.

Electrically polarized BEC can interact with the beam of charged
and polarized particles by means of charge-dipole and
dipole-dipole interaction. Such interaction leads to transfer of
energy from beam to medium and, consequently, to generation of
waves. In plasma physics the effect of generation of waves by
 electron ~\cite{Bret PRE 04} or magnetized neutron
~\cite{Andreev PIERS 2011}  beam are well-known. In presented
paper we consider similar effect in polarized BEC.

We start from the equations of quantum hydrodynamics. System of
QHD equations consists of continuity equation, momentum balance
equation, eq. of polarization evolution and eq. of polarization
current evolution. We made first principles derivation of this
equation. For this purpose we used manyparticle Schr\"{o}dinger
equation. In this article we analytically calculate the dispersion
properties of polarized BEC. We have shown that the dynamics of
EDM leads to existence of new branch in dispersion law.
Consequently, there are the waves of polarization in polarized
BEC, along with Bogoliubov's mode. We obtain appropriate
contribution to the dispersion of Bogoliubov's mode related to
DDI. Then, we consider the process of wave generation in polarized
BEC by means of monoenergetic beam of neutral polarized particles.
We predict new method of generation of Bogoliubov's mode and a new
type of modes, rest of the polarization waves.

This paper is organized as follows. We introduce the model
Hamiltonian in Sec. II, and present the  momentum balance
equation. Farther, in Sec. II, we derive the equation of evolution
of polarization and obtain the influence of the interactions on
the evolution of polarization. In Sec. III  we calculate the
dispersion dependence of EE in BEC. The polarization evolution is
taken into account. We obtain the contribution of polarization in
dispersion of Bogoliubov's mode and show the existence of new wave
solution. In Sec. IV we study the wave generation in polarized BEC
by means of neutral polarized particle beam. In Sec. V we present
the brief summary of our results.

\section{\label{sec:level1}II. The model}

Explicit form of the hamiltonian of considering system in a
quasi-static approximation is

$$\hat{H}=\sum_{i}\Biggl(\frac{1}{2m_{i}}\hat{\textbf{p}}_{i}^{2}-d_{i}^{\alpha}E_{i,ext}^{\alpha}\Biggr)$$
\begin{equation}\label{di BEC Hamiltonian}+\frac{1}{2}\sum_{i,j\neq i}\Biggl(U_{ij}-d_{i}^{\alpha}d_{j}^{\beta}G_{ij}^{\alpha\beta}\Biggr).\end{equation}
The first term here is the operator for kinetic energy. The second
term represents the interaction between the dipole moment
$d_{i}^{\alpha}$ and the external electrical field. The subsequent
terms represent short-range $U_{ij}$ and dipole-dipole
interactions between particles, respectively. The Green's function
for dipole-dipole interaction is taken as
$G_{ij}^{\alpha\beta}=\nabla^{\alpha}_{i}\nabla^{\beta}_{i}(1/r_{ij})$.

For investigation of EE dynamic in polarized BEC we derive the
system of QHD equations. This system of equations consists of the
continuity equation, the Euler's equation and for the case of
polarized particles the equations of polarization evolution and
equation of field. The system of equations is derived by methods
described in ~\cite{Andreev PRA08}.

The first equation of a QHD equations system is the continuity
equation
\begin{equation}\label{di BEC cont eq}\partial_{t}n(\textbf{r},t)+\partial^{\alpha}(n(\textbf{r},t)v^{\alpha}(\textbf{r},t))=0.\end{equation}

The momentum balance equation for the polarized BEC has the form
$$mn(\textbf{r},t)(\partial_{t}+\textbf{v}\nabla)v^{\alpha}(\textbf{r},t)+\partial_{\beta}p^{\alpha\beta}(\textbf{r},t)$$
$$-\frac{\hbar^{2}}{4m}\partial^{\alpha}\triangle
n(\textbf{r},t)+\frac{\hbar^{2}}{4m}\partial^{\beta}\Biggl(\frac{\partial^{\alpha}n(\textbf{r},t)\cdot\partial^{\beta}n(\textbf{r},t)}{n(\textbf{r},t)}\Biggr)
$$
$$=n(\textbf{r},t)\Upsilon\partial^{\alpha}n(\textbf{r},t)+\frac{1}{2}\Upsilon_{2}\partial^{\alpha}\triangle n^{2}(\textbf{r},t)$$
\begin{equation}\label{di BEC bal imp eq short}+P^{\beta}(\textbf{r},t)\partial^{\alpha}E^{\beta}(\textbf{r},t),
\end{equation}
where
\begin{equation}\label{di BEC Upsilon} \Upsilon=\frac{4\pi}{3}\int
dr(r)^{3}\frac{\partial U(r)}{\partial r},
\end{equation}
and
 \begin{equation}\label{di BEC Upsilon2}\Upsilon_{2}\equiv\frac{\pi}{30}\int dr
(r)^{5}\frac{\partial U(r)}{\partial r}.\end{equation}

In equation (\ref{di BEC bal imp eq short})  we defined a
parameter $\Upsilon_{2}$ as (\ref{di BEC Upsilon2}). This
definition differs from the one in the work ~\cite{Andreev PRA08}.
Terms proportional to $\hbar^{2}$ appear as a result of usage of
quantum kinematics. The first two members at the right side of the
equation (\ref{di BEC bal imp eq short}) are first terms of
expansion of the quantum stress tensor. They occur because of
taking into account of the SRI potential $U_{ij}$. The interaction
potential $U_{ij}$ determinates the macroscopic interaction
constants $\Upsilon$ and $\Upsilon_{2}$. The last two members of
the equation (\ref{di BEC bal imp eq short}) describe force fields
that affect the dipole moment in a unit of volume as the effect of
the external electrical field and the field produced by other
dipoles, respectively. The last member is written using the
self-consistent field approximation ~\cite{MaksimovTMP 1999}.
$p^{\alpha\beta}(\textbf{r},t)$ is a tensor of the kinetic
pressure, which depends on particles' thermal velocities and does
not contribute into the BEC dynamics at temperatures near zero.

 Also, we have a field equation
\begin{equation}\label{di BEC field eq}\nabla\textbf{E}(\textbf{r},t)=-4\pi \nabla\textbf{P}(\textbf{r},t).\end{equation}

In the case particles do not bear the dipole moment, the
continuity equation and the momentum balance equation form a
closed system of equations. When the dipole moment is taken into
account in a momentum balance equation, a new physical value
emerges, a polarization vector field $P^{\alpha}(\textbf{r},t)$.
This causes system of equations to become incomplete.

Next equation we need for investigation of EE
dispersion is the equation of polarization evolution
\begin{equation}\label{di BEC eq polarization}\partial_{t}P^{\alpha}(\textbf{r},t)+\partial^{\beta}R^{\alpha\beta}(\textbf{r},t)=0.\end{equation}

The equation (\ref{di BEC eq polarization}) does not contain
information about the effect of the interaction on the
polarization evolution. The evolution equation of
$R^{\alpha\beta}(\textbf{r},t)$ can be constructed by analogy with
the above derived equations. Method of the equations derivation is
described in Appendix. Using a self-consistent field approximation
of the dipole-dipole interaction we obtain an equation for the
polarization current $R^{\alpha\beta}(\textbf{r},t)$ evolution
 $$\partial_{t}R^{\alpha\beta}(\textbf{r},t)+\partial^{\gamma}\biggl(R^{\alpha\beta}(\textbf{r},t)v^{\gamma}(\textbf{r},t)$$
$$+R^{\alpha\gamma}(\textbf{r},t)v^{\beta}(\textbf{r},t)-P^{\alpha}(\textbf{r},t)v^{\beta}(\textbf{r},t)v^{\gamma}(\textbf{r},t)\biggr)$$
$$+\frac{1}{m}\partial^{\gamma}r^{\alpha\beta\gamma}(\textbf{r},t)-\frac{\hbar^{2}}{4m^{2}}\partial_{\beta}\triangle P^{\alpha}(\textbf{r},t)$$
$$+\frac{\hbar^{2}}{8m^{2}}\partial^{\gamma}\biggl(\frac{\partial_{\beta}P^{\alpha}(\textbf{r},t)\partial_{\gamma}n(\textbf{r},t)}{n(\textbf{r},t)}+\frac{\partial_{\gamma}P^{\alpha}(\textbf{r},t)\partial_{\beta}n(\textbf{r},t)}{n(\textbf{r},t)}\biggr)
$$
$$=\frac{1}{m}\Upsilon\partial^{\beta}(n(\textbf{r},t)P^{\alpha}(\textbf{r},t))$$
\begin{equation}\label{di BEC eq for pol current gen selfconsist
appr}+\frac{1}{m}\frac{P^{\alpha}(\textbf{r},t)P^{\gamma}(\textbf{r},t)}{n(\textbf{r},t)}\partial^{\beta}E^{\gamma}(\textbf{r},t).\end{equation}
Here  $r^{\alpha\beta\gamma}(\textbf{r},t)$ represents the
contribution of thermal movement of polarized particles into the
dynamics of $R^{\alpha\beta}(\textbf{r},t)$. As we deal with BEC
below, the contribution of $r^{\alpha\beta\gamma}(\textbf{r},t)$
may be neglected. The last term of the formula (\ref{di BEC eq for
pol current gen selfconsist appr}) includes both external
electrical field and a self-consistent field that particle dipoles
create. Used in (\ref{di BEC eq for pol current gen selfconsist
appr}) approximations are described in Appendix.

The first term in right side of Eq. (\ref{di BEC eq for pol
current gen selfconsist appr}) describe the short range
interaction.

We can see various interactions are included in the equations
(\ref{di BEC bal imp eq short}) and (\ref{di BEC eq for pol
current gen selfconsist appr}) additively. At a short distances
among particles acts both SRI and dipole-dipole interaction. At
large distances remain only dipole-dipole interaction.

From Eq. (\ref{di BEC eq for pol current gen selfconsist appr}) we
can see that the change of polarization arise from both the
dipole-dipole interaction and the short range interaction. The SRI
among particle leads to displacement of particles. Consequently,
as particles has EDM, where are motion of EDM, e.i. changing of
the $R^{\alpha\beta}(\textbf{r},t)$.

The terms that are proportional to $\hbar^{2}$ are of quantum
origin as they are analogs of the Bohm quantum potential in the
momentum balance equilibrium.

\section{\label{sec:level1}III. Elementary excitations in the polarized BEC}

We can analyze the linear dynamics of elemental excitations in the
polarized BEC using the QHD equations (\ref{di BEC cont eq}),
(\ref{di BEC bal imp eq short}), (\ref{di BEC field eq}), (\ref{di
BEC eq polarization}) and (\ref{di BEC eq for pol current gen
selfconsist appr}). Let's assume the system is placed in an
external electrical field $\textbf{E}_{0}=E_{0}\textbf{e}_{z}$.
The values of concentration $n_{0}$ and polarization
$\textbf{P}_{0}=\kappa\textbf{E}_{0}$ for the system in an
equilibrium state are constant and uniform and its velocity field
$v^{\alpha}(\textbf{r},t)$ and tensor
$R^{\alpha\beta}(\textbf{r},t)$ values are zero.

We consider the small perturbation of equilibrium state like
$$\begin{array}{ccc}n=n_{0}+\delta n,& v^{\alpha}=0+v^{\alpha},& \end{array}$$
\begin{equation}\label{di BEC equlib state BEC}\begin{array}{ccc}& & P^{\alpha}=P_{0}^{\alpha}+\delta P^{\alpha}, R^{\alpha\beta}=0+\delta R^{\alpha\beta}.\end{array}\end{equation}
Substituting these relations into system of equations (\ref{di BEC
cont eq}), (\ref{di BEC bal imp eq short}), (\ref{di BEC eq
polarization}), (\ref{di BEC eq for pol current gen selfconsist
appr}) and (\ref{di BEC field eq}) \textit{and} neglecting
nonlinear terms, we obtain a system of linear homogeneous
equations in partial derivatives with constant coefficients.
Passing to the following representation for small perturbations
$\delta f$
$$\delta f =f(\omega, \textbf{k}) exp(-\imath\omega+\imath \textbf{k}\textbf{r}) $$
 yields the homogeneous system of algebraic equations.
The electric field strength is assumed to have a nonzero value.
Expressing all the quantities entering the system of equations in
terms of the electric field, we come to the equation
$$\Lambda\cdot E_{z}=0,$$
where
$$\Lambda=\omega^{2}-\frac{\hbar^{2}k^{4}}{4m^{2}}+\frac{\Upsilon k^{2}n_{0}}{2m}+4\pi\sigma\frac{P_{0}^{2}k^{2}}{mn_{0}}$$
$$-\frac{2\pi\Upsilon k^{4}P_{0}^{2}}{m^{2}\omega^{2}-\hbar^{2}k^{4}/4+m\Upsilon k^{2}n_{0}-m\Upsilon_{2}k^{4}n_{0}}.$$ In this case, the dispersion equation is
$$\Lambda=0.$$
Solving this equation with respect to $\omega^{2}$ we obtain a
following results.

The dispersion characteristic for EE in BEC
can be expressed in the form of
$$\omega^{2}=\frac{1}{2m}\Biggl(-\frac{3}{2}\Upsilon n_{0}k^{2}+\frac{\hbar^{2}k^{4}}{2m}+\Upsilon_{2}n_{0}k^{4}+ 4\pi\sigma\frac{P_{0}^{2}k^{2}}{n_{0}}$$
\begin{equation}\label{di BEC general disp dep}\pm\sqrt{\biggl(\frac{1}{2}\Upsilon n_{0}k^{2}-\Upsilon_{2}n_{0}k^{4}+ 4\pi\sigma\frac{P_{0}^{2}k^{2}}{n_{0}}\biggr)^{2}-8\pi\Upsilon k^{4}P_{0}^{2}}\Biggr)\end{equation}

In contrast to the nonpolarized BEC, where only Bogoliubiv's mode
exists ~\cite{Bogoliubov N.1947,L.P.Pitaevskii RMP 99,Andreev
PRA08}, a new wave solution appears in a polarized system due to
the polarization dynamics. Bogoliubov's mode corresponds to a
negative solution of the equation (\ref{di BEC general disp dep}).
 New wave solution it is a wave of polarization. In general case
presented by formula (\ref{di BEC general disp dep}) the frequency
of polarization wave in BEC depend on $P_{0}$ and $\Upsilon$. To
investigate the BEC polarization effect on the Bogoliubov's mode
and the dispersion characteristic of the new solution we analyze
extreme cases of the formula (\ref{di BEC general disp dep}).

Formula (\ref{di BEC general disp dep}) demonstrates that taking account of the BEC
polarization dynamics leads to a new solution.

Let's start with the case when the effect of polarization is low
compared to the contribution of short-range effects, i.e. of
members proportional to $\Upsilon$ and $\Upsilon_{2}$. If so, then
formula (\ref{di BEC general disp dep}) takes the form
$$\omega^{2}_{B}=\frac{\hbar^{2}k^{4}}{4m^{2}}-\frac{\Upsilon n_{0}k^{2}}{m}+\frac{\Upsilon_{2}n_{0}k^{4}}{m}$$
 \begin{equation}\label{di BEC 01a}-\frac{4\pi P_{0}^{2}k^{2}}{mn_{0}}\frac{\Upsilon(\sigma-1)-2\sigma\Upsilon_{2}k^{2}}{\Upsilon-2\Upsilon_{2}k^{2}},\end{equation}
$$\omega^{2}_{P}=\frac{\hbar^{2}k^{4}}{4m^{2}}-\frac{1}{2}\frac{\Upsilon
n_{0}k^{2}}{m}$$
\begin{equation}\label{di BEC 01b}+\frac{4\pi P_{0}^{2}k^{2}}{mn_{0}}\frac{\Upsilon(\sigma-1)-2\sigma\Upsilon_{2}k^{2}}{\Upsilon-2\Upsilon_{2}k^{2}}.\end{equation}
Here we use indexes "B" for Bogoliubov's mode and "P" for new
polarization mode. When deriving formulae (\ref{di BEC 01a}) and
(\ref{di BEC 01b}) we expanded a sub-radical expression in
(\ref{di BEC general disp dep}) and took only first two terms of
the expansion to estimate the influence of polarization on the
wave dispersion.

As it follows from equations (\ref{di BEC eq polarization}) and
(\ref{di BEC eq for pol current gen selfconsist appr}), changes in
polarization can occur due dipole-dipole interactions as well as
to SRI and to quantum Bohm potential, i.e. to members proportional
to $\hbar^{2}$. That's the reason for existence of other ways of
polarization change when the contribution of equilibrium
polarization $P_{0}$ is neglectable. In the latter case, taken in
a linear approximation, changes in the polarization do not affect
the concentration evolution.

Formulae (\ref{di BEC general disp dep}) are valid even in the
absence of the external electrical field when equilibrium
polarization equals zero $P_{0}=0$. Equations (\ref{di BEC general
disp dep}) in that case take the form
\begin{equation}\label{di BEC disp in absence ext field a}\omega^{2}_{B}=\frac{1}{m}\Biggl(\frac{\hbar^{2}k^{4}}{4m}-\Upsilon n_{0}k^{2}+\Upsilon_{2}n_{0}k^{4}\Biggr),\end{equation}
\begin{equation}\label{di BEC disp in absence ext field b}\omega^{2}_{P}=\frac{1}{2m}\Biggl(\frac{\hbar^{2}k^{4}}{2m}-\Upsilon n_{0}k^{2}\Biggr).\end{equation}

The waves of polarization could be existed  at the absence of
external electric field $E_{0}=0$. In this case the equilibrium
state is no polarized and dipole direction of particles is
distributed accidentally.

If the contribution of equilibrium polarization into BEC
dispersion (\ref{di BEC general disp dep}) is comparable to the
contribution of SRI in the third order of the
interaction radius  $\Upsilon_{2}$ relationships (\ref{di BEC
general disp dep}) are transformed into
\begin{equation}\label{di BEC 02a}\omega^{2}_{B}=\frac{\hbar^{2}k^{4}}{4m^{2}}-\frac{\Upsilon n_{0}k^{2}}{m}+\frac{\Upsilon_{2}n_{0}k^{4}}{m}+4\pi\frac{P_{0}^{2}k^{2}}{mn_{0}},\end{equation}
\begin{equation}\label{di BEC 02b}\omega^{2}_{P}=\frac{\hbar^{2}k^{4}}{4m^{2}}-\frac{1}{2}\frac{\Upsilon n_{0}k^{2}}{m}+4\pi\frac{P_{0}^{2}k^{2}}{mn_{0}}(\sigma-1).\end{equation}

Using Feshbach's resonance ~\cite{Feshbach resonance} we can
transform the SRI potential in such a way that $\Upsilon=0$ while
$\Upsilon_{2}\neq 0$. Formulas (\ref{di BEC general disp dep}) in
this situation turn into
\begin{equation}\label{di BEC 03a}\omega^{2}_{B}=\frac{\hbar^{2}k^{4}}{4m^{2}}+\frac{\Upsilon_{2}n_{0}k^{4}}{m},\end{equation}
\begin{equation}\label{di BEC 03b}\omega^{2}_{P}=\frac{\hbar^{2}k^{4}}{4m^{2}}+4\pi\sigma\frac{P_{0}^{2}k^{2}}{mn_{0}}.\end{equation}

In this paper we primarily focus on the influence of BEC
polarization on its dispersion characteristics. So, let's consider
the case when the contribution into the dispersion of the
SRI at the first order of the interaction
radius, i.e. terms proportional to $\Upsilon$, is comparable to the
contribution of polarization, and their total effect is much
greater than the contribution of terms proportional to
$\Upsilon_{2}$. Here, (\ref{di BEC general disp dep}) turns into
\begin{equation}\label{di BEC 05a}\omega^{2}_{B}=\frac{\hbar^{2}k^{4}}{4m^{2}}-\frac{\Upsilon n_{0}k^{2}}{m}\biggl(\frac{(\sigma-1)}{\sigma}\biggr),\end{equation}
$$\omega^{2}_{P}=\frac{\hbar^{2}k^{4}}{4m^{2}}+4\pi\sigma\frac{P_{0}^{2}k^{2}}{mn_{0}}$$
\begin{equation}\label{di BEC 05b}-\frac{1}{2}\frac{\Upsilon n_{0}k^{2}}{m}\biggl(\frac{(\sigma+4)}{2\sigma}\biggr).\end{equation}

The the first order of the interaction radius interaction constant
for dilute gases has the form

$$\Upsilon=-\frac{4\pi\hbar^{2}a}{m},$$
where $a$ is the scattering length SL ~\cite{L.P.Pitaevskii RMP
99,Andreev PRA08}. The value $\Upsilon_{2}$ may be expressed
approximately as
$$\Upsilon_{2}=-\frac{\theta a^{2}\Upsilon}{8},$$
where $\theta$ is a constant positive value about 1, which depends
on the interatomic interaction potential. Finally, $\Upsilon_{2}$
takes the form
\begin{equation}\label{di BEC Upsilon 2 approx}\Upsilon_{2}=-\frac{\pi\theta\hbar^{2}a^{3}}{2m}.\end{equation}

\section{\label{sec:level1}IV. Generation of waves in polarized BEC}

In that section we consider the process of wave generation in BEC
by means of the beam of neutral polarized particles. The
interaction between beam and BEC has dipole-dipole origin.

 To get the dispersion solution  we use the system of QHD
equations for each sort of particles (\ref{di BEC cont eq}),
(\ref{di BEC bal imp eq short}), (\ref{di BEC eq polarization}),
(\ref{di BEC eq for pol current gen selfconsist appr}) and the
equation of field (\ref{di BEC field eq}). The equilibrium state
of system is characterized by following values of the BEC
parameters:
$$\begin{array}{ccc}n=n_{0}+\delta n,& v^{\alpha}=0+v^{\alpha},& \end{array}
$$
\begin{equation}\label{di BEC equlib state BEC}\begin{array}{ccc}& & P^{\alpha}=P_{0}^{\alpha}+\delta P^{\alpha}, R^{\alpha\beta}=0+\delta R^{\alpha\beta}\end{array}
\end{equation}
and values of the beam parameters:
$$\begin{array}{ccc}n_{b}=n_{0b}+\delta n_{b},& v^{\alpha}_{b}=U\delta^{z\alpha}+\delta v^{\alpha}_{b},& \end{array}
$$
\begin{equation}\label{di BEC equlib state beam}\begin{array}{ccc} & P_{b}^{\alpha}=P_{0b}^{\alpha}+\delta P^{\alpha}_{b},& R^{\alpha\beta}_{b}=R^{\alpha\beta}_{0b}+\delta R^{\alpha\beta}_{b}.\end{array}
\end{equation}
The polarization $P_{0}^{\alpha}$ is proportional to external
electric field $E_{0}^{\alpha}$. We consider the case then
$\textbf{E}_{0}=[E_{0}sin\varphi, 0, E_{0}cos\varphi]$. In this
case the tensor $R^{\alpha\beta}_{0b}$ has only two unequal to
zero elements: $R^{zx}_{0b}=R_{0b}sin\varphi$ and
$R^{zz}_{0b}=R_{0b}cos\varphi$.
 For the process, under consideration the dispersion relation is:
 $$1+4\pi k^{2}\Biggl(\frac{P_{0}^{2}}{\omega^{2}-\frac{\hbar^{2}k^{4}}{4m^{2}}+\frac{\Upsilon n_{0}k^{2}}{2m}}\times$$
 $$\times\biggl(\frac{\Upsilon k^{2}/(2m^{2})}{\omega^{2}-\frac{\hbar^{2}k^{4}}{4m^{2}}+\frac{\Upsilon n_{0}k^{2}}{m}-\frac{\Upsilon_{2}n_{0}k^{4}}{m}}-\frac{\sigma}{mn_{0}}\biggr) $$
$$+\frac{1}{(\omega-k_{z}U)^{2}-\frac{\hbar^{2}k^{4}}{4m^{2}_{b}}}\times$$
 \begin{equation}\label{di BEC disp dependence with the beam} \times\biggl(\frac{2(\omega-k_{z}U)P_{0b}k_{z}(P_{0b}U-R_{0b})}{m_{b}n_{0b}\biggl((\omega-k_{z}U)^{2}-\frac{\hbar^{2}k^{4}}{4m^{2}_{b}}\biggr)}-\frac{\sigma_{b}P_{0b}^{2}}{m_{b}n_{0b}}\biggr)\Biggr)=0.\end{equation}
Using relation $P_{0b}U-R_{0b}=0$, we can simplify the equation
(\ref{di BEC disp dependence with the beam}) and obtain

$$1+\frac{\omega_{D}^{2}}{\omega^{2}-\omega_{1}^{2}}\Biggl(\frac{\Upsilon n_{0}k^{2}/(2m)}{\omega^{2}-\omega_{2}^{2}}-\sigma\Biggr)$$
\begin{equation}\label{di BEC disp dependence with the beam simpl}-\frac{\sigma_{b}\omega_{Db}^{2}}{(\omega-k_{z}U)^{2}-\frac{\hbar^{2}k^{4}}{4m^{2}_{b}}}=0\end{equation}
In this formula the following designations are used
\begin{equation}\label{di BEC omega 1}\omega_{1}^{2}=\frac{\hbar^{2}k^{4}}{4m^{2}}-\frac{\Upsilon n_{0}k^{2}}{2m},\end{equation}
\begin{equation}\label{di BEC omega 2}\omega_{2}^{2}=\frac{\hbar^{2}k^{4}}{4m^{2}}-\frac{\Upsilon n_{0}k^{2}}{m}+\frac{\Upsilon_{2}n_{0}k^{4}}{m}\end{equation}
and
\begin{equation}\label{di BEC omega D}\omega_{Di}^{2}=\frac{4\pi P_{0i}^{2}k^{2}}{m_{i}n_{0i}},\end{equation}
where $i$ is the index of sorts of particles, the BEC or the beam.

The equation (\ref{di BEC disp dependence with the beam simpl})
has two beam related solutions, in the absence of BEC medium:
\begin{equation}\label{di BEC beam mode 1}\omega=k_{z}U\pm\sqrt{\frac{\hbar^{2}k^{4}}{4m^{2}_{b}}+\sigma_{b}\omega_{Db}^{2}}.\end{equation}
We will consider the possibilities of instabilities for the case
of low-density beam, the limit case $\omega_{Db}\sim
n_{0b}\rightarrow 0$. In this case we can neglect the last term in
square root in (\ref{di BEC beam mode 1}).  The resonance
interaction beam with the BEC realizing if
\begin{equation}k_{z}U\pm\frac{\hbar k^{2}}{2m_{b}}=\omega(k),\end{equation}
and could lead to instabilities. The quantity $\omega(k)$ is the
dispersion of BEC modes (\ref{di BEC general disp dep}).  The
frequency in this case can be presented in the form
\begin{equation}\label{di BEC influence of beam mode}\omega=k_{z}U\pm\frac{\hbar k^{2}}{2m_{b}}+\delta\omega.\end{equation}
Lets consider two limit cases.

\subsection{small frequency shift limit}

In the limit case
\begin{equation}\label{di
BEC}\delta\omega\ll\hbar k^{2}/m\end{equation}
the frequency shift
obtain in the form:
\begin{equation}\label{di BEC increment} \delta\omega^{2}=\pm\frac{2\sigma_{b}m_{b}m^{2}\omega_{Db}^{2}(\omega^{2}-\omega_{1}^{2})^{2}(\omega^{2}-\omega_{2}^{2})^{2}}{\omega\omega_{D}^{2}\hbar\Upsilon^{2}n_{0}^{2}k^{6} W}, \end{equation}
where
\begin{equation}\label{di BEC W general}W=2\omega^{2}-\omega_{1}^{2}-\omega_{2}^{2}-\frac{2m\sigma}{\Upsilon n_{0}k^{2}}(\omega^{2}-\omega_{2}^{2})^{2}.\end{equation}
and the frequency $\omega$ determined with formula (\ref{di BEC
general disp dep}). The instabilities take place in the case
$\delta\omega^{2}<0$. The sign of $\delta\omega^{2}$ is depend on
the sign of $W$.

For the case resonance interaction of beam with the waves in BEC
there are instabilities, for the first beam mode in (\ref{di BEC
beam mode 1}) at $W<0$ and for the second beam mode in formula
(\ref{di BEC beam mode 1}) if $W>0$. For the polarization mode $W$
is positive. It means that the interaction of polarization mode
with the second beam related mode results in the instability. For
the Bogoliubov's mode the sign of $W$ depend on $\sigma$.

We can consider the following cases:

(i) the contribution of equilibrium polarization to $\omega(k)$ is
dominant; then $W>0$;

(ii) the dominant contribution to  $\omega(k)$ results from the
term in  $\omega(k)$ which is proportional to $\Upsilon$,
\textit{and}, equilibrium polarization and SPI in TOIR give
comparable contribution to $\omega(k)$. In this case there is
$$\sigma_{0}=1+\frac{\Upsilon_{2}n_{0}k^{4}/m}{\Upsilon
n_{0}k^{2}/(2m)+2\Upsilon_{2}n_{0}k^{4}/m-4\omega_{D}^{2}}$$. The
sign of $W$ varies at $\sigma=\sigma_{0}$. The dependence of $W$
from $\sigma$ is presented in the table 1.

\begin{table}[b]\caption{\label{tab:table1}%
In this table the sign of the $W$ are presented for the
Bogoliubov's mode when SRI is prevailed.}
\begin{ruledtabular}
\begin{tabular}{lcr}
 \textrm{ }&
 \textrm{\large $\Upsilon>0$}&
\multicolumn{1}{c}{\textrm{\large $\Upsilon<0$}}\\
\colrule
 & (attraction) & (repulsion)\\
  &  &  \\
\large
$\sigma>\sigma_{0}$ & \large + & \large -\\
 &  &  \\
\large $\sigma<\sigma_{0}$ & \large - & \large +\\
\end{tabular}
\end{ruledtabular}
\end{table}

\subsection{large frequency shift limit}

In the limit case
\begin{equation}\label{di
BEC large limit}\delta\omega\gg\hbar k^{2}/m\end{equation} we have

\begin{equation}\label{di BEC W rep in pol wave Q}\delta\omega  =\xi\sqrt[3]{\frac{\sigma_{b}m^{2}\omega_{Db}^{2}(\omega^{2}-\omega_{1}^{2})^{2}(\omega^{2}-\omega_{2}^{2})^{2}}{\omega\omega_{D}^{2}\Upsilon^{2}n_{0}^{2}k^{4}\mid W\mid}},\end{equation}
where $\xi$ equal to $\xi_{1}=\sqrt[3]{1}$ for $W>0$ or
$\xi_{-1}=\sqrt[3]{-1}$ for $W<0$. Evident form of quantities
$\xi_{1}$ and $\xi_{-1}$ is
$$\xi_{-1}=[-1, \frac{-1+\imath\sqrt{3}}{2}, \frac{-1-\imath\sqrt{3}}{2}]$$
 and
$$\xi_{1}=[1, \frac{1+\imath\sqrt{3}}{2}, \frac{1-\imath\sqrt{3}}{2}].$$

The considerations concerning the sign of $W$, reflected in the
table 1, are valid also for the limit condition (\ref{di BEC large
limit}).

From the formulas (\ref{di BEC increment}) and (\ref{di BEC W rep
in pol wave Q}) we can see that external neutral particles beam
leads to instabilities for both Bogoliubovs and polarization
waves.

\section{\label{sec:level1}V. Conclusion}

In this work we developed the method for description of dynamics
of polarized BEC. This method accounts for the effect of
polarization on changes in the concentration and in velocity
field, which are determined in general by the continuity equation
and Euler's equation. We derived the evolution equations of the
polarization and the polarization current. The equations derived
contain information about the influence of the interactions on the
polarization evolution. We studied the effect of polarization on
the BEC dynamics and the influence of SRI on the polarization
evolution. An expression of SRI contribution in the equation of
polarization current evolution via concentration, polarization and
the SRI potential $\Upsilon=\Upsilon(U_{ij})$ was derived. With
the assumption that the state of polarized Bose particles in the
form of BEC can be described with some single-particle wave
function. Changes in polarization due to SRI are shown to be
determined at the first order of the interaction radius by the
same interaction constant that occurs in Euler's equation and
Gross-Pitaevskii equation.

The dispersion of EE in the polarized BEC was analyzed. In the
article, we show that polarization evolution in BEC causes a novel
type of waves in BEC. The effect of polarization on the dispersion
of the Bogoliubov's mode and the dispersion of a new wave mode
were studied.

We show the possibility of wave generation in polarized BEC by
means of monoenergetic beam of neutral polarized particles.

\section{\label{sec:level1}Appendix}

At derivation of system of QHD equations we need to write evident
form of Hamiltonian of dipole-dipole interaction. In some works
the dynamics of the magnetic dipole moment and of the EDM
~\cite{Lahaye RPP 09,Yi PRA 02} are analyzed in similar
ways. Usual expressions of a Hamiltonian for dipole-dipole
interaction are equal for electric and magnetic dipoles:

$$H_{dd}=\frac{\delta^{\alpha\beta}-3r^{\alpha}r^{\beta}/r^{2}}{r^{3}}d_{1}^{\alpha}d_{2}^{\beta}.$$

However it has been shown by Breit ~\cite{spin-spin interaction}
that a Hamiltonian for spin-spin interaction, and, as a
consequence, for the interaction of magnetic moments contains a
term that is proportional to Dirac $\delta$-function
$\delta(r_{1}-r_{2})d_{1}^{\alpha}d_{2}^{\beta}$. The coefficient
of the $\delta$-function has been refined later ~\cite{MaksimovTMP
2001} so that the Hamiltonian is in accord with Maxwell's free
equations, such as $div\textbf{B}=0$. The resultant expression for
the spin-spin interaction Hamiltonian is:
$$H_{\mu\mu}=\Biggl(4\pi\delta_{\alpha\beta}\delta(\textbf{r}_{12})+\nabla^{\alpha}_{1}\nabla^{\beta}_{1}(1/r_{12})\Biggr)\mu^{\alpha}_{1}\mu^{\beta}_{2}.$$

Thus, the differences in the dipole-dipole interactions of
electric dipoles, and magnetic dipoles, must be taken into account
in the development of theoretical field apparatus.

The Schr\"{o}dinger equation defines wave function in a
3N-dimensional configuration space. Physical processes in systems
that involve large number of bodies occur in a three-dimensional
physical space ~\cite{Goldstein}. That's why a problem evolves of
obtaining a quantum-mechanical description of a system of
particles in terms of material fields e.g. concentration, momentum
density, energy density and other fields of various tensor
dimension that are defined in a three-dimension space.

In equations (\ref{di BEC bal imp eq short}), (\ref{di BEC field
eq}) polarization occurs in the form of
\begin{equation}\label{di BEC def polarization}P^{\alpha}(\textbf{r},t)=\int dR\sum_{i}\delta(\textbf{r}-\textbf{r}_{i})\psi^{*}(R,t)\hat{d}_{i}^{\alpha}\psi(R,t),\end{equation}
where $\textbf{r}_{i}$ is the coordinate operator for i-th particle, $dR=\prod_{p=1}^{N}d\textbf{r}_{p}$.

In right side of equation (\ref{di BEC eq for pol current gen
selfconsist appr}) where is the force-like field
$F^{\alpha\beta}(\textbf{r},t)$ which give rise to evolution of
the polarization current $R^{\alpha\beta}(\textbf{r},t)$. In
general form, for $F^{\alpha\beta}(\textbf{r},t)$, we can write:
 \begin{equation}\label{di BEC }F^{\alpha\beta}(\textbf{r},t)=-\frac{1}{m}\partial_{\gamma}\Sigma^{\alpha\beta\gamma}(\textbf{r},t)+\frac{1}{m}D^{\alpha\gamma}(\textbf{r},t)\partial^{\beta}E^{\gamma}(\textbf{r},t).\end{equation}

A tensor
 \begin{equation}\label{di BEC }D^{\alpha\beta}(\textbf{r},t)=\int dR\sum_{i}\delta(\textbf{r}-\textbf{r}_{i})d_{i}^{\alpha}d_{i}^{\beta}\psi^{*}(R,t)\psi(R,t),\end{equation}
occurs in the term that represents a dipole-dipole interaction and
an interaction of the dipole with external electrical field. Based
on the reasons of dimensions this value can be approximately
presented as
\begin{equation}\label{di BEC appr for D}D^{\alpha\beta}(\textbf{r},t)=\sigma\frac{P^{\alpha}(\textbf{r},t)P^{\beta}(\textbf{r},t)}{n(\textbf{r},t)}.\end{equation}

A SRI causes the tensor $\Sigma^{\alpha\beta\gamma}(\textbf{r},t)$ to occur in the equation (\ref{di BEC eq for pol current gen selfconsist appr}). Taken at the first order of the interaction radius it has the form
$$\Sigma^{\alpha\beta\gamma}(\textbf{r},t)=-\frac{1}{2}\int dR\sum_{i,j\neq i}\delta(\textbf{r}-\textbf{R}_{ij})$$
\begin{equation}\label{di BEC short range in polariz}\times\frac{r_{ij}^{\beta}r_{ij}^{\gamma}}{r_{ij}}\frac{\partial U_{ij}}{\partial r_{ij}}\psi^{*}(R,t)\hat{d}_{i}^{\alpha}\psi(R,t).\end{equation}
Tensor $\Sigma^{\alpha\beta\gamma}(\textbf{r},t)$ describe the
influence of SRI on evolution of polarization.
Eq. (\ref{di BEC short range in polariz}) describe the SRI.

If we apply the procedure described in ~\cite{Andreev PRA08} to
the calculation of the quantum stress tensor, and neglect the
contribution of thermal excitations,
$\Sigma^{\alpha\beta\gamma}(\textbf{r},t)$ for BEC takes a form of
\begin{equation}\label{di BEC 3ind-Sigma}\Sigma^{\alpha\beta\gamma}_{BEC}(\textbf{r},t)=-\frac{1}{2}\Upsilon\delta^{\beta\gamma}n(\textbf{r},t)P^{\alpha}(\textbf{r},t).\end{equation}

Formula (\ref{di BEC 3ind-Sigma}) is obtained for the case where
particles located in state with the lowest energy, which could be
described by one particle wave function. This state may be the
product of strong interaction.

Tensor $\Sigma^{\alpha\beta\gamma}(\textbf{r},t)$ is, therefore,
like the quantum stress tensor
$\sigma^{\alpha\beta}(\textbf{r},t)$ in the momentum balance
equation (\ref{di BEC bal imp eq short}), dependent on $\Upsilon$
at the first order of the interaction radius ~\cite{Andreev PRA08}.


\end{document}